\documentclass[conference]{IEEEtran}

\usepackage[most]{tcolorbox}
\usepackage{algorithmic}
\usepackage{textcomp}
\usepackage{xcolor}
\usepackage{booktabs}
\usepackage{array}
\usepackage{graphicx}
\usepackage{siunitx}
\usepackage{multirow}
\usepackage{comment}
\usepackage[hidelinks]{hyperref}

\usepackage{color}

\usepackage{tabularx}
\def\BibTeX{{\rm B\kern-.05em{\sc i\kern-.025em b}\kern-.08em
    T\kern-.1667em\lower.7ex\hbox{E}\kern-.125emX}}

\begin{document}

\title{\textbf{TAM-Eval}: Evaluating LLMs for Automated Unit Test Maintenance}

\author{

\IEEEauthorblockN{
Elena Bruches\IEEEauthorrefmark{4}\IEEEauthorrefmark{1},
Vadim Alperovich\IEEEauthorrefmark{2}\IEEEauthorrefmark{1},
Dari Baturova\IEEEauthorrefmark{4},
Roman Derunets\IEEEauthorrefmark{4},
Daniil Grebenkin\IEEEauthorrefmark{4},
Georgy Mkrtchyan\IEEEauthorrefmark{2}
}

\IEEEauthorblockN{
Oleg Sedukhin\IEEEauthorrefmark{4},
Mikhail Klementev\IEEEauthorrefmark{4},
Ivan Bondarenko\IEEEauthorrefmark{3},
Nikolay Bushkov\IEEEauthorrefmark{2},
Stanislav Moiseev\IEEEauthorrefmark{2}
}

\IEEEauthorblockA{
\IEEEauthorrefmark{4}\textit{Siberian Neuronets LLC}, Novosibirsk, Russia\\
\IEEEauthorrefmark{2}\textit{T-Technologies}, Moscow, Russia\\
\IEEEauthorrefmark{3}\textit{Novosibirsk State University}, Novosibirsk, Russia\\
\{bruches, baturova, derunets, grebenkin, sedukhin, klementev\}@sibnn.ai\\
\{v.alperovich, g.p.mkrtchyan, n.bushkov, s.moiseev\}@t-tech.dev\\
i.bondarenko@g.nsu.ru 
}

\IEEEauthorblockA{\IEEEauthorrefmark{1} Both authors contributed equally to this research.}
}

\maketitle

\begin{abstract}
While Large Language Models (LLMs) have shown promise in software engineering, their application to unit testing remains largely confined to isolated test generation or oracle prediction, neglecting the broader challenge of test suite maintenance.

We introduce \textbf{TAM-Eval} (\textbf{T}est \textbf{A}utomated \textbf{M}aintenance Evaluation), a framework and benchmark designed to evaluate model performance across three core test maintenance scenarios: creation, repair, and updating of test suites. Unlike prior work limited to function-level tasks, TAM-Eval operates at the test file level, while maintaining access to full repository context during isolated evaluation, better reflecting real-world maintenance workflows.

Our benchmark comprises 1,539 automatically extracted and validated scenarios from Python, Java, and Go projects. TAM-Eval supports system-agnostic evaluation of LLMs, using a reference-free protocol based on test suite pass rate, code coverage, and mutation testing.

Empirical results indicate that state-of-the-art LLMs have limited capabilities in realistic test maintenance processes and yield only marginal improvements in test effectiveness. We release TAM-Eval as an open-source framework to support future research in automated software testing. Our data and code are available at \url{https://github.com/trndcenter/TAM-Eval}.
\end{abstract}


\section{Introduction}
\label{sec:intro}
Unit testing plays a pivotal role in ensuring robustness and reliability by validating individual components of software~\cite{10.5555/1349795}, with industry reports estimating that testing consumes up to 25\% of total project budgets~\cite{globalapptestingMuchDoes}.

Recent advances in large language models (LLMs) have significantly impacted software engineering tasks, including code generation~\cite{jiang2024surveylargelanguagemodels}, test generation~\cite{yang2024evaluationlargelanguagemodels}, documentation~\cite{dvivedi2024analysis}, refactoring ~\cite{cordeiro2024empiricalstudycoderefactoring}, and bug fixing ~\cite{meng2024empiricalstudyllmbasedagents}. Despite strong performance in function-level test generation, existing LLM-based tools largely produce isolated tests or assertions and do not address the broader lifecycle of real-world test suites.

In practice, however, software quality depends heavily on test maintenance — the continuous generation, repair, and updating of tests as code evolves. Failing to maintain tests leads to outdated suites, broken pipelines, and undetected regressions. Yet, this setting remains underexplored in current research.

\textbf{Technical Challenges.} Evaluating automated test maintenance requires context-aware reasoning over both production code and existing tests, as well as dynamic execution environments that support correctness checking and incremental evaluation across dimensions such as execution correctness, coverage and mutation score.

\begin{figure*}[!t]
    \centering
    \includegraphics[width=0.75\textwidth]{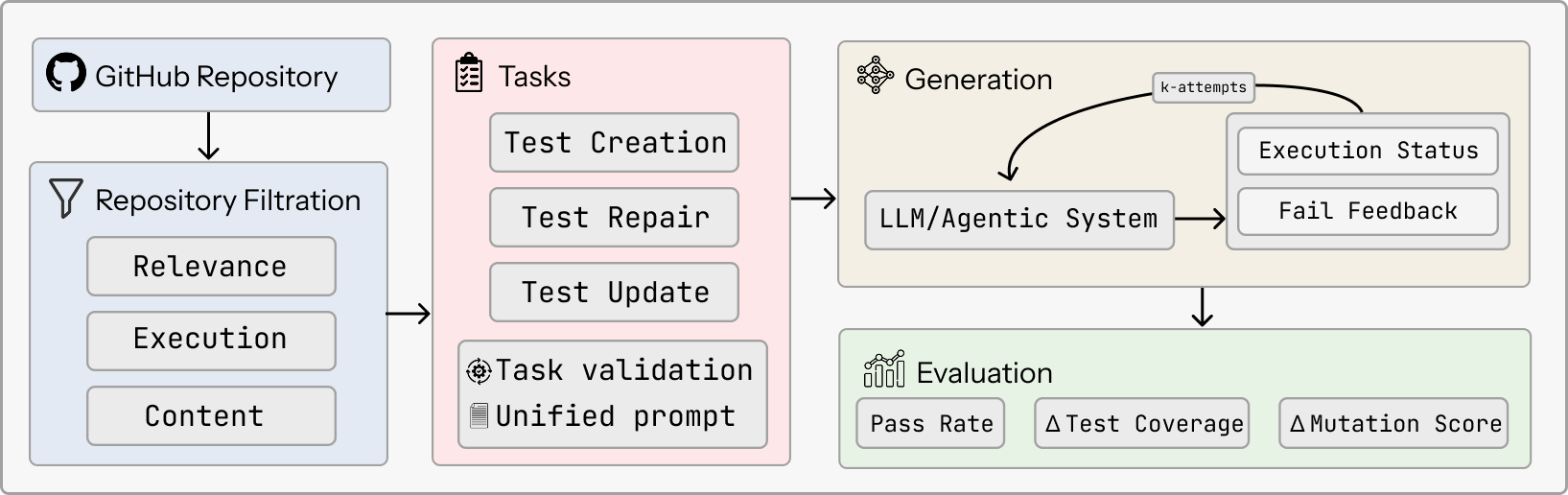}
    \caption{Overview of TAM-Eval, a framework and benchmark for evaluating LLMs on unit test maintenance. The pipeline filters GitHub repositories, constructs tasks for test creation, repair, and update with execution-based validation, performs LLM-driven test generation with iterative feedback, and computes Pass Rate, $\Delta$Test Coverage, and $\Delta$Mutation Score.}
    \label{fig:main}
\end{figure*}

\textbf{Our Contributions.} We present \textbf{TAM-Eval} (\textbf{T}est \textbf{A}utomated \textbf{M}aintenance \textbf{E}valuation), a benchmark and evaluation framework for comprehensive unit test maintenance assessment. An overview of its pipeline is shown in Fig.~\ref{fig:main} Its key contributions are as follows:

\begin{itemize}
    \item \textbf{Test Maintenance Scenarios:} The novelty of TAM-Eval is that it addresses the complete test maintenance cycle --- \emph{generation}, \emph{repair}, and \emph{update} of test suites at the granularity of entire test files, reflecting realistic developer workflows.

    \item \textbf{Curated Real-world Dataset:} The benchmark includes 1,539 validated scenarios from actively maintained open-source repositories, which were selected via a multi-stage filtering pipeline to ensure quality with measures to minimize data contamination.

    \item \textbf{Unified and Feedback-Aware Evaluation:} All tasks are prompted through a unified prompt template with minimal instructions and iterative execution feedback, testing the model's ability to generalize across maintenance types and simulate practical workflows. 

    \item \textbf{Extensible, Open Evaluation Suite:} TAM-Eval supports reference-free evaluation via test pass rate, coverage, and mutation testing. It is language-agnostic, easy to extend, and fully open-sourced to facilitate reproducibility and community benchmarking.

    \item \textbf{Empirical Analysis of LLMs:} We evaluate advanced SOTA (state-of-the-art) LLMs, revealing that even the best-performing models exhibit large variability across tasks and languages, with mutation coverage improvements rarely exceeding $12$ percentage points, indicating substantial room for advancement. This highlights the need for developing more context-aware methods and distilling supervision from verifiers such as compilers, mutation, and coverage signals.
\end{itemize}

\textbf{Outline.} Section~\ref{sec:background} reviews relevant work. Section~\ref{sec:bench} presents our benchmark design. Section~\ref{sec:framework} details the evaluation framework. Section~\ref{sec:experimental_setup} describes the experimental setup. Section~\ref{sec:results} reports our experimental results. 

\section{Background}
\label{sec:background}
\subsection{Unit tests generation}

 The attempts to automate the unit testing process started in the last century \cite{10.1109/TSE.1976.233818} and focused on code coverage, such as Randoop \cite{10.1145/1297846.1297902} or EvoSuite \cite{10.1145/2025113.2025179}.

With the rise of LLMs, various prompting techniques \cite{gao2025promptalchemistautomatedllmtailored} were used to generate unit tests. Whereas earlier models had difficulties with self-repair and test improvement \cite{huang2024largelanguagemodelsselfcorrect,olausson2024repair}, such methods have received a second wind with the advent of reasoning approaches \cite{10.5555/3600270.3602070,10.5555/3666122.3666639}. For example, ChatTester \cite{10.1145/3660783} leverages ChatGPT itself to improve the quality of its generated tests. \cite{gao2025promptalchemistautomatedllmtailored} Libro \cite{10.1109/ICSE48619.2023.00194} focuses on post-processing steps that help discern when LLMs are effective and rank the produced tests according to their validity. 

Rapid development in code generation and particularly in test generation has led to the need to create high-quality assessment systems for different test-related tasks like test repairing or test updating.

\subsection{Unit tests repairing}

One of the important tasks associated with unit tests is the task of test repair, a process that helps validate the code and improve its performance. 

The frameworks and methods like TestART \cite{gu2025testartimprovingllmbasedunit}, ChatUniTest \cite{10.1145/3663529.3663801}, already mentioned ChatTester \cite{10.1145/3660783} and others \cite{gu2025llmtestgenerationiterative,10329992} use the APR (Automated Program Repair) idea, but the proposed pipelines also contain the repair stage, which can be rule-based, LLM-based, or combined. The rule-based repair strategy can include steps like syntax analysis and searching test annotations for test extraction, copying imports from focal file (tested file) to test file \cite{10.1145/3663529.3663801}, mutations of different entities to increase test correctness and robustness \cite{gu2025testartimprovingllmbasedunit}. However, some research shows that in spite of the fact that repair can be a part of the test generation pipeline, these tasks require a different set of LLM skills \cite{NEURIPS2024_94f093b4}. Nevertheless, the use of different LLM strategies and prompt engineering allows the creation of architectures like TestGen-LLM \cite{10.1145/3663529.3663839}, which was fine-tuned to increase test’s coverage and reliability.  

\subsection{Unit tests updating}

Unit tests need to be periodically updated and maintained: they may not initially cover all the necessary functionality or be unstable to mutations, for which purpose systems are developed that can improve existing tests, such as Meta's TestGen-LLM tool \cite{10.1145/3663529.3663839}. 

The solution of this issue may be associated with using commits that modify the tests. 
Application of the corresponding concept can be found in the dataset with vulnerabilities of open-source Android projects, collected on the basis of commits with vulnerabilities and fix-commits \cite{10.1145/3508398.3511495}. The GitBug-Actions benchmark also used a buggy commit and a subsequent fixing commit.  Long Code Arena \cite{bogomolov2024longcodearenaset} used commit data for CI Builds Repair tasks as well as for Project-Level Code Completion. 

Noteworthy is SWE-smith \cite{yang2025swesmith}, where to create a training set the authors took a current commit and brought its state to a tests failure. In one strategy, they reverted pull requests on commits back to the point where the tests started failing. SWT-Bench~\cite{NEURIPS2024_94f093b4} extends this paradigm by explicitly pairing GitHub issues with their corresponding fix commits, requiring generated tests to fail on pre-fix code while passing on post-fix implementations.

\subsection{Unit-testing related benchmarks}

Unlike traditional code-evaluation benchmarks focused mainly on assessing code-generation abilities of large language models (LLMs), specialized benchmarks geared toward evaluating unit-testing capabilities have recently emerged. Each benchmark diverges in the specific tasks that it evaluates.

For instance, TestBench \cite{zhang2024testbench} concentrates on Java programs, consisting of 108 meticulously curated examples drawn from nine large-scale open-source projects. It evaluates LLMs across five key dimensions: syntactic correctness, compilation success, test validity, code coverage, and defect-detection rates. 
Another notable benchmark is ProjectTests \cite{wang2025projecttestprojectlevelllmunit}, which spans three major programming languages (Python, Java, and JavaScript) and incorporates 20 exemplary projects per language. These projects underwent extensive manual selection and refinement, ensuring the dataset's reliability. However, the high cost of expanding this dataset might hinder its long-term utility.
SWT-Bench ~\cite{NEURIPS2024_94f093b4}, derived from SWE-Bench ~\cite{jimenez2024swebench}, boasts over 2,000 Python samples, showcasing diverse application contexts. Meanwhile, CPP-UT-Bench \cite{bhargava2024cpputbenchllmswritecomplex} compiles 2,653 manually constructed code-unittest pairs from 14 renowned open-source C++ repositories. Notably, its cross-domain representation enhances the relevance to real-world problems.
Additionally, CLOVER \cite{xu2025clovertestcasegeneration} addresses Python-specific unit test generation by treating the task as both code completion and open-ended code generation.

Some benchmarks delve into more sophisticated goals beyond simple test-case generation. For example, TESTEVAL \cite{wang-etal-2025-testeval} examines three distinct dimensions: overall code coverage (cov@k), targeted line and branch coverage, and path coverage. However, its absence of repository-level data restricts its practical relevance.

In addition, several benchmarks emulate real-world scenarios. DevBench ~\cite{li2024promptinglargelanguagemodels} simulates the entire software development lifecycle (SDLC), integrating tasks such as software design, environment setup, implementation, acceptance testing, and unit testing. Similarly, TestGenEval ~\cite{jain2025testgenevalrealworldunit} prioritizes the generation of comprehensive test suites, stressing initial test authorship, suite expansion, and coverage improvement. 

In contrast with the benchmarks mentioned above, we propose a fully automated pipeline to collect high-quality data for unit test creation, test repair, and test updating tasks.

\section{Benchmark Construction}
\label{sec:bench}
We provide a benchmark for evaluation of automated unit test maintenance process, which contains 1,539 samples for 3 programming languages: Java, Python, and Go. These samples were meticulously filtered and annotated with supplementary metadata to approximate real-world scenarios. The basic information about the benchmark is presented in Table~\ref{tab1}.

Below, we describe the benchmark construction process in detail.

\subsection*{\textbf{Stage 1. Repository Selection}}

We collected open-source GitHub repositories via GitHub API that satisfy quality and relevance criteria: updated after \texttt{2020}, and the focal files used in our benchmark to have their latest commit after \texttt{2025} (\texttt{2024} for Java), at least 40 stars, permissive licenses (MIT, Apache-2.0), and at least two contributors. We expect that these criteria help avoid LLM training contamination and can be updated over time.

After this stage, 94.1\% of the original data was filtered out.

\subsection*{\textbf{Stage 2. Execution-based Filtering}}

Since our framework executes code during evaluation, it is crucial to ensure that candidate projects are both buildable and runnable within reasonable resource and time constraints. To this end, we apply the following filters:

\textbf{Automated build process.} Only projects that were constructed entirely through predefined commands, without any manual intervention, were included.

\textbf{Test execution.} We require that the test suite runs within 30 seconds.

\textbf{Test stability.} To eliminate flaky or non-deterministic tests, each test is executed twice. Only those that pass consistently across both runs are retained. At this step, it became also possible to distinguish suitable focal-test file pairs through the analysis of test file imports and filenames, according to the features of every chosen programming language. The test files could contain only imports from focal file, both of the files in a suite could not have imports from non-standard packages, etc.

\textbf{Test coverage level.} We exclude focal-test file pairs where the original test suite achieves less than 40\% line coverage of the corresponding focal function. This ensures that retained test suites are not trivially under-specified and that there is a meaningful baseline from which to assess improvements in coverage and mutation score.

After applying these filters, we retain only those repositories that contain at least \textbf{five focal-test file pairs} that satisfy the above conditions. 
An additional 12.8\% of the data present at the beginning of the stage was filtered out.

\subsection*{\textbf{Stage 3. Content-based Filtering}}

To ensure that the dataset comprises relevant and high-quality samples suitable for robust evaluation, we applied a series of content-based checks:

\textbf{Minimum number of test cases.} 
Samples containing fewer than two test cases in the associated test file were excluded.

\textbf{Function complexity in focal files.} 
Pairs where the focal file does not contain at least one function with five or more lines of executable code were removed. This excludes trivial or stub functions unlikely to provide meaningful evaluation challenges.

\textbf{File size constraints.} 
Pairs where focal or test files contain fewer than 20 lines or exceed the 99th percentile were discarded. 

\textbf{Comment density.} 
Samples in which comments made up over 70\% of the focal file were excluded, ensuring that the model receives primarily executable content. We believe that a high ratio of comments may cause the data leakage and help models to generate tests based on the documentation but not on the code itself.

\textbf{Generated code.}
Automatically generated files (e.g., containing "Generated by" markers) were removed to avoid non-human-authored samples.

Finally, to prevent over-representation of large repositories and to promote diversity in the benchmark, we performed balanced sampling across projects. Specifically, we randomly sampled up to 10 focal–test file pairs per repository. 

After the third stage, 45\% of the data remaining from the previous stage was filtered out.

\subsection*{\textbf{Stage 4. Test Maintenance Tasks Creation}}

In the final stage, we derive concrete evaluation tasks from the filtered and executable focal–test file pairs, the samples were assigned to tasks with no overlap. These tasks are designed to reflect core subproblems within the domain of automated test maintenance. 






\textbf{A. Test Creation Task} -- generating new tests for previously untested or insufficiently tested code. 
We simulate three distinct test creation scenarios by applying structured modifications to the collected dataset:

\textbf{From Scratch:} we cleared the contents of the test file corresponding to the focal file, effectively removing the entire test suite.  

\textbf{Add New Tests:} selecting test files with partial coverage of the focal file (i.e. coverage less than 100\%), requiring extension to cover uncovered logic.

\textbf{Recover Tests:} removing a subset of test cases from test files with high coverage, thereby creating the need to generate missing test logic while maintaining overall coverage. The described setup guarantees the feasibility of new test generation. In contrast, lower coverage in the \textit{Add New Test} setting does not always imply that additional test cases are necessary, useful, or even feasible.

\textbf{B. Test Repair Task} -- correcting broken, outdated, or inconsistent tests. To construct this task samples, two sources of faulty tests were implemented: (1) synthetically injected errors via heuristics based on the \texttt{tree-sitter}\footnote{\url{https://tree-sitter.github.io/tree-sitter/}}, and (2) LLM-generated defects prompted to produce diverse error types and complexities, grounded in both human-written and synthetic test suites. Specifically, Qwen2.5 Coder 32B was used to generate broken test code.
 
Below we provide the description of the defects which were incorporated into the test suites.

\textbf{Syntax errors} (4.07\% of repair challenge):  
basic syntactic violations such as missing braces, commas, or extraneous tokens. This category ensures that models can perform minimal syntax correction

\textbf{Execution-targeted faults} (47.37\% of repair challenge): 
syntactically valid tests that fail at compile or runtime due to missing imports, undefined identifiers, invalid calls, or exceptions. Introduced via \texttt{tree-sitter}-based heuristics and LLM prompting, and validated through execution logs.

\textbf{Coverage-targeted breakdowns} (17.77\% of repair challenge):  
tests that compile and run but add little or no value to code coverage or mutation score, often due to redundant logic or superficial variation.

\textbf{Efficiency-targeted breakdowns} (30.74\% of repair challenge): tests lacking assertions or checks, thus offering weak behavioral guarantees. Simulated by removing verification logic to assess the model’s ability to restore meaningful oracles.

\textbf{C. Test Updating Task} -- adapting existing tests to reflect changes in the associated production code. 
To simulate a realistic test maintenance process, we reverted the test file to its state $k$ commits earlier while keeping the current focal file. This ensures that test updates are both necessary and feasible, as evidenced by degradation in test coverage, mutation score, or execution correctness.

We identify suitable commits by traversing the test file’s history in reverse chronological order and selecting the first commit that modifies at least one line of executable code. 
A sample is retained if test or mutation coverage drops by at least 5\% compared to the current version, or if the reverted test fails to execute correctly with the current focal file.

\begin{table}[!t]
\caption{Benchmark statistics}
\label{tab1}
\centering
\small
\renewcommand{\arraystretch}{1.2}
\begin{tabularx}{\columnwidth}{ l *{7}{>{\centering\arraybackslash}X} }
\toprule
\textbf{Lang} & \textbf{\# samples} & \textbf{\# repos} & \textbf{Avg. focal LOC} & \textbf{Avg. test LOC} & \textbf{\# test cases} & \textbf{Avg. TestCov (\%)} & \textbf{Avg. MutCov (\%)} \\
\midrule
Python & 442 & 55 & 151 & 89 & 5.59 & 50.40 & 52.27 \\
Java   & 495 & 55 & 96  & 66 & 3.89 & 32.45 & 27.50 \\
Go     & 602 & 82 & 119 & 114 & 3.22 & 31.70 & 43.24 \\
\midrule
Overall & 1539 & 192 & 120 & 91 & 4.11 & 37.32 & 17.81 \\
\bottomrule
\end{tabularx}
\end{table}

\section{Evaluation Framework}
\label{sec:framework}
To assess LLM-driven systems for automated unit test maintenance, we introduce a reference-free and fully automated evaluation pipeline. The pipeline supports all three constructed tasks introduced in this work: test creation, repair, and updating.

\subsection{Metrics}

We define a set of interpretable test-oriented metrics that capture the effectiveness of generated test suites in improving coverage, detecting faults, and ensuring correctness.

\textbf{PassRate.}  
A basic yet essential metric: The ratio of successful test executions to total executed tests. It serves as a filter for utility, as failing tests cannot reliably validate correctness. Similar execution-based correctness signals have been used previously, e.g., in \textit{MERA Code} via the \textit{CodeCorrectness} benchmark~\cite{chervyakov2025meracodeunifiedframework}.

\textbf{\(\Delta\)TestCov.}  
Line-level test coverage reflects how thoroughly the test suite exercises the focal file. For a given focal-test file pair, we define:
\[
  \texttt{TestCov} = \frac{\text{\# lines executed by tests}}{\text{\# total executable lines in the focal file}} \times 100
\]
We compute the coverage both before $\texttt{TestCov}_{\text{initial}}$ and after $\texttt{TestCov}_{\text{final}}$ modifications introduced by an evaluated system. The relative improvement is:
\[
  \Delta\texttt{TestCov} = \texttt{TestCov}_{\text{final}} - \texttt{TestCov}_{\text{initial}}
\]

\textbf{\(\Delta\)MutCov.}  
Mutation testing evaluates whether a test suite can distinguish the correct implementation from faulty variants. Let \(F\) be the original file under test. We generate mutants \(F^{mut}_1, \dots, F^{mut}_k\) by applying small code changes (e.g., negating a condition or altering an operator). An ideal test suite should: pass on \(F\) (the unmodified code), fail on as many mutants \(F^{mut}_i\) as possible ("failed" or "killed" mutants ). The number of generated mutants depends on the structure of the target code.

Therefore, the formula for mutation coverage is:
\[
  \texttt{MutCov} = \frac{\text{\# failed mutants}}{\text{\# total mutants}} \times 100,
\]

And subsequently:
\[
  \Delta\texttt{MutCov} = \texttt{MutCov}_{\text{final}} - \texttt{MutCov}_{\text{initial}}
\]

\subsection{Evaluation Pipeline}

Each benchmark sample undergoes a reproducible and fully automated evaluation process consisting of the following steps:

\textbf{Sanity Checks:} The generated test file is validated for non-triviality -- we discard outputs that are empty or exact duplicates of the input.

\textbf{Syntax Validation:} Language-specific parsers (based on \texttt{tree-sitter}) check that the generated code is syntactically valid and compilable.

\textbf{Test Execution and Coverage Analysis:} The test suite is executed against the corresponding focal file. We collect line-level test coverage using language-specific tools to compute \(\Delta\)TestCov.

\textbf{Mutation Testing:} We generate a set of mutants and evaluate how many are failed by the test suite to compute \(\Delta\)MutCov. For each run, the set of mutants is fixed, which means that e.g. mutpy (for Python) generates all possible mutants for the focal file without any sampling. 

\textbf{Metric Aggregation:} Individual deltas are computed for each metric and then aggregated across the benchmark to produce average performance indicators.

\subsection{Infrastructure}

To ensure consistency, scalability, and isolation, all evaluations are conducted in sandboxed Docker environments. For each project, we automatically detect appropriate build and test commands using static heuristics based on language conventions and project structure. 

We use the following tools for execution and analysis:

\textbf{Coverage analysis:} \texttt{coverage.py}\footnote{\url{https://github.com/nedbat/coveragepy}} for Python, \texttt{JaCoCo}\footnote{\url{https://github.com/jacoco/jacoco}} for Java, and \texttt{cover package} \footnote{\url{https://pkg.go.dev/cmd/cover}} for Go.

\textbf{Mutation testing:} \texttt{mutpy}\footnote{\url{https://github.com/mutpy/mutpy}} for Python, \texttt{PIT}\footnote{\url{https://pitest.org}} for Java, and \texttt{go-mutesting}\footnote{\url{https://github.com/avito-tech/go-mutesting}} for Go.

Our infrastructure is modular and language-extensible, allowing straightforward addition of new languages or testing tools in future iterations of the framework.

\section{Experimental Setup}
\label{sec:experimental_setup}
\begin{figure*}[t]
    \centering
    
    \includegraphics[width=0.9\textwidth]{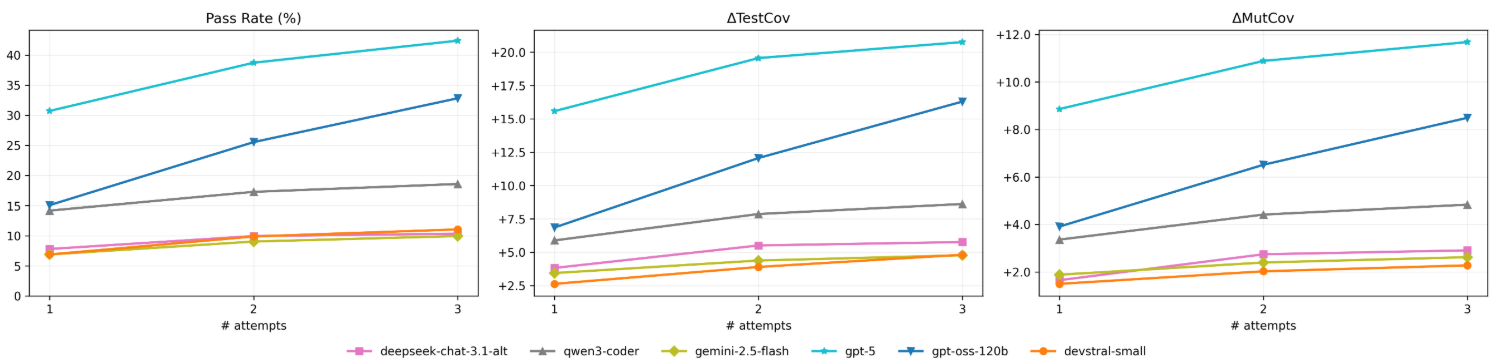}
    \caption{Dynamics across three metrics: Pass Rate, mean $\Delta$ Test Coverage, and mean $\Delta$ Mutation Coverage by attempts}
    \label{fig:metrics-main}
\end{figure*}

\subsection{Generation stage}
The model receives as input the focal file and its corresponding test file contents. While our default setup limits the input context to streamline and accelerate the evaluation process, the framework allows full-repository context use. The model is prompted to rewrite the entire test file from scratch to enhance the provided test suite by increasing test coverage and effectiveness.

\textbf{Failure Recovery via Attempts.}
Each model is allowed up to $k$ \textit{attempts} to generate a valid output. If the initial generation fails, the next attempt includes failure feedback -- such as syntax errors, compiler messages, or stack traces -- automatically extracted with minimal preprocessing. 

\textbf{Unified Prompt Instruction.}
All test maintenance tasks -- generation, repair, and update -- share a single unified instruction template. The prompt provides minimal task-specific guidance, relying on the model's capability to infer intent from context. 

\subsection{Inference Configuration}

We evaluate a diverse set of LLMs, both open-source and proprietary, to benchmark their performance on automated unit test maintenance. The list of models is presented in Table~\ref{tab:main_results}.

We queried all models via the OpenRouter API\footnote{\url{https://openrouter.ai}} using the chat/completions endpoint through the AsyncOpenAI client. Sampling was performed with a fixed random seed and temperature of $0.25$ to ensure reproducibility while allowing minimal stochasticity. Default system prompts were used unless stated otherwise; for Qwen3 Coder 480B A35B we enabled the \texttt{/no\_think} option to suppress verbose reasoning and encourage direct edits.

\section{Results}
\label{sec:results}
We present the evaluation results of \textsc{TAM-Eval} across various model configurations and task dimensions. 

\begin{table}[!t]
\caption{Main Results (attempt@3)}
\label{tab:main_results}
\centering
\small
\setlength{\tabcolsep}{3pt}
\renewcommand{\arraystretch}{1.05}
\resizebox{\columnwidth}{!}{%
\begin{tabular}{l c c c}
\toprule
\textbf{Model} &
\textbf{Pass Rate (\%)} &
\textbf{$\overline{\Delta}$ TestCov} &
\textbf{$\overline{\Delta}$ MutCov} \\
\midrule
Devstral-Small \cite{mistral2025devstral}            & 11.04 & $\uparrow 4.8 \pm 0.5$ & $\uparrow 2.3 \pm 0.4$ \\
Qwen3 Coder 480B A35B \cite{yang2025qwen3technicalreport}      & 18.58 & $\uparrow 8.6 \pm 0.6$ & $\uparrow 4.8 \pm 0.5$ \\
DeepSeek V3.1 671B \cite{deepseekai2024deepseekv3technicalreport}       & 10.33 & $\uparrow 5.8 \pm 0.5$ & $\uparrow 2.9 \pm 0.4$ \\
GPT-OSS-120B  \cite{openai2025gptoss120bgptoss20bmodel}             & \textbf{32.81} & $\uparrow \textbf{16.3} \pm 0.8$ & $\uparrow \textbf{8.5} \pm 0.7$ \\
\midrule
Gemini 2.5 Flash \cite{2025arXiv250706261C}        & 9.94 & $\uparrow 4.8 \pm 0.5$ & $\uparrow 2.6 \pm 0.4$ \\
GPT-5 \cite{openai2025gpt41}                      & \textbf{42.37} & $\uparrow \textbf{20.8} \pm 1.0$ & $\uparrow \textbf{11.7} \pm 0.8$ \\
\bottomrule
\end{tabular}}
\end{table}

\subsection{Main Results}
Table~\ref{tab:main_results} summarizes the overall performance of the evaluated models with a maximum of three recovery attempts (Attempt@3). GPT-5 achieves the highest $PassRate$ of 42.3\%, significantly outperforming other models, highlighting its superior ability to generate effective test cases. GPT-OSS-120B follows with a $PassRate$ of 32.8\%, the largest Qwen's model Qwen3 Coder 480B A35B demonstrates moderate performance as well as DeepSeek V3.1 671B. 

We also analyze the impact of multiple attempts on model performance, as shown in Fig.~\ref{fig:metrics-main}. Across all models, all three metrics generally improve with additional attempts, with GPT-5 and GPT-OSS-120B showing the most consistent gains, underscoring their robustness in iterative recovery abilities and utilizing fail feedback signals.

However, current models demonstrate limited capability in test maintenance tasks on their first attempt without fail feedback context, as evidenced by the low initial performance in Fig.~\ref{fig:metrics-main}, except GPT-5, which achieves a $PassRate$ of 30.7\% at Attempt@1.

\subsection{Performance by Language}

Table~\ref{tab:main_results_lang} presents models' performance by programming language at Attempt@3. GPT-5 consistently outperforms all other models, achieving the highest average \textit{Pass Rate} across languages, most notably dominating Go and Java. GPT-5 also shows the largest mean improvements across languages having \(\overline{\Delta}\)TestCov 18.7 and \(\overline{\Delta}\)MutCov 10.2). Although GPT-OSS-120B is the second-best model overall, it achieves the highest Python line-level and mutation test coverage. Qwen3 Coder 480B A35B provides decent metrics, delivering particularly strong Go results.

Surprisingly, Go has emerged as the most suitable language for modern LLMs, thanks to its concise syntax, strict typing, and minimal semantic noise — making code generation more accurate, predictable, and maintainable. Java, although it sometimes gives a fairly high \textit{Pass Rate}
 (e.g., 29.3\% for GPT-5), shows notably lower \(\overline{\Delta}\)TestCov and \(\overline{\Delta}\)MutCov, highlighting that test executability and passing do not always translate into meaningful improvements in coverage for more verbose, statically-typed languages.

Across all evaluated languages, Qwen3 Coder 480B A35B consistently produces the shortest code snippets, whereas Gemini 2.5 Flash outputs the longest. Specifically, for Python, the mean test file lengths range from 3,698 characters (Qwen3 Coder 480B A35B) to 15,641 characters (Gemini 2.5 Flash), for Go: from 4,410 (Qwen3 Coder 480B A35B) to 13,463 (Gemini 2.5 Flash), and for Java: from 3,017 (Qwen3 Coder 480B A35B) to 12,711 (Gemini 2.5 Flash).

Interestingly, despite having shorter overall test files, Java exhibits the highest assert density. Models produce an average of 15 asserts per file (Qwen3 Coder 480B A35B) up to 66 asserts (Gemini 2.5 Flash). By contrast, Python averages between 12 (Qwen3 Coder 480B A35B) and 63 (Gemini 2.5 Flash) asserts, while Go shows significantly fewer assertions, ranging from just 10 (Qwen3 Coder 480B A35B) to 30 (Gemini 2.5 Flash). 

Additionally, the distribution of test cases mirrors the assert count trends: Python typically contains between 9 (Qwen3 Coder 480B A35B) and 19 (Devstral-Small) test cases per file, Go ranges from 7 (Qwen3 Coder 480B A35B, GPT-OSS-120B, Gemini 2.5 Flash) to 11 (Devstral-Small), and Java spans from 10 (GPT-OSS-120B) to 16 (Devstral-Small).

\begin{table*}[!t]
  \caption{Main results by language (attempt@3)}
  \label{tab:main_results_lang}
  \centering
  \small
  \setlength{\tabcolsep}{3pt}
  \resizebox{\textwidth}{!}{%
  \begin{tabular}{
      @{} l c c c | c c c | c c c @{}
  }
  \toprule
  \multirow{2}{*}{\textbf{Model}} & \multicolumn{3}{c}{\textbf{Go}} & \multicolumn{3}{c}{\textbf{Java}} & \multicolumn{3}{c}{\textbf{Python}} \\
  \cmidrule(r){2-4} \cmidrule(r){5-7} \cmidrule(r){8-10}
     & \textbf{Pass Rate (\%)} & \textbf{$\overline{\Delta}$ TestCov} & \textbf{$\overline{\Delta}$ MutCov} &
       \textbf{Pass Rate (\%)} & \textbf{$\overline{\Delta}$ TestCov} & \textbf{$\overline{\Delta}$ MutCov} &
       \textbf{Pass Rate (\%)} & \textbf{$\overline{\Delta}$ TestCov} & \textbf{$\overline{\Delta}$ MutCov} \\
  \midrule
  Devstral-Small        & $\uparrow$21.1 & $\uparrow$10.7 & $\uparrow$6.6  & $\uparrow$3.4 & $\uparrow$0.2 & $\downarrow$0.0 & $\uparrow$5.9 & $\uparrow$2.0  & $\downarrow$0.7 \\
  Qwen3 Coder 480B A35B & $\uparrow$34.4 & $\uparrow$17.7 & $\uparrow$12.1 & $\uparrow$6.3 & $\uparrow$1.1 & $\uparrow$0.2 & $\uparrow$10.9 & $\uparrow$4.7  & $\uparrow$0.6 \\
  DeepSeek V3.1 671B & $\uparrow$18.9 & $\uparrow$11.9 & $\uparrow$7.6  & $\uparrow$4.6 & $\uparrow$1.7 & $\downarrow$0.0 & $\uparrow$5.0 & $\uparrow$1.9  & $\downarrow$0.1 \\
  GPT-OSS-120B         & $\uparrow$\textbf{56.5} & $\uparrow$\textbf{30.9} & $\uparrow$\textbf{21.7} & $\uparrow$\textbf{14.3} & $\uparrow$\textbf{4.2} & $\uparrow$\textbf{0.6} & $\uparrow$\textbf{21.3} & $\uparrow$\textbf{10.0} & $\uparrow$\textbf{1.1} \\
  \midrule
  Gemini 2.5 Flash      & $\uparrow$17.1 & $\uparrow$9.7  & $\uparrow$6.3  & $\uparrow$6.1 & $\uparrow$1.3 & $\downarrow$0.0 & $\uparrow$4.5 & $\uparrow$1.9  & $\uparrow$\textbf{0.6}   \\
  GPT-5                 & $\uparrow$\textbf{67.8} & $\uparrow$\textbf{38.8} & $\uparrow$\textbf{28.1} & $\uparrow$\textbf{29.3} & $\uparrow$\textbf{8.1} & $\uparrow$\textbf{1.9} & $\uparrow$\textbf{20.4} & $\uparrow$\textbf{9.2}  & $\uparrow$0.5  \\       
  \bottomrule
  \end{tabular}}
\end{table*}

\subsection{Performance by Task}
 While GPT-5 remains the overall top performer, the gap between models varies by task. Create and Repair tasks generally yield higher \textit{Pass Rates} and coverage gains, suggesting that generating or fixing tests aligns well with model capabilities. Update task, in contrast, remains challenging: all models show reduced \(\overline{\Delta}\)MutCov, reflecting the difficulty of precise context-aware edits. Interestingly, GPT-OSS-120B achieves competitive results in Update, approaching GPT-5 in \textit{Pass Rate} and \(\overline{\Delta}\)TestCov. This may reflect better alignment with structured modification patterns.

Within the test creation task, models achieve the highest \(\overline{\Delta}\)TestCov and \(\overline{\Delta}\)MutCov in the \textit{From Scratch} scenario, where entire test suites must be synthesized. This is partly due to the fact that coverage metrics in this setting start from zero — making any successful generation yield large absolute gains. In contrast, improving already partially tested code, as in \textit{Add New Tests} or \textit{Recover Tests}, proves harder: the higher the initial coverage, the more challenging it becomes to find meaningful, non-redundant testing paths. Remarkably, \textit{Add New Tests} and \textit{Recover Tests} show relatively high \textit{Pass Rates} but negligible coverage improvements, indicating limited semantic depth in model completions. Tables with models' performance by task and scenario are provided in our repository~\footnote{https://github.com/trndcenter/TAM-Eval}.

\subsection{Failures Analysis}

\begin{figure}[htbp]
\includegraphics[width=\columnwidth]{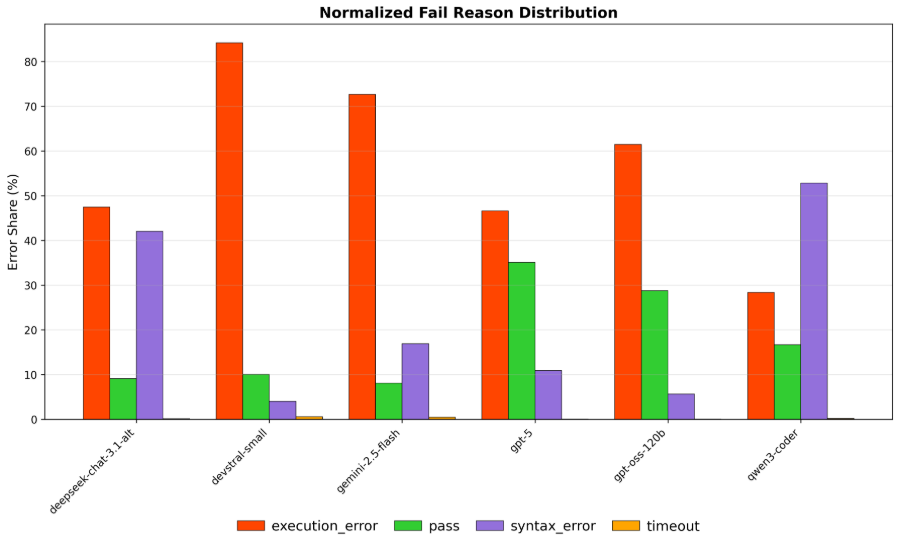}
\caption{Distribution of generated test suites fail reason (attempt@3).}
\label{fig:fail_reasons}
\end{figure}

Fig.~\ref{fig:fail_reasons} reveals that the dominant failure mode across all models is \texttt{execution\_error}, accounting for over 60\% of invalid outputs for most of the evaluated LLMs. This indicates persistent challenges in maintaining runtime-correct behavior, even when syntactic validity is achieved. Notably, models like GPT-OSS-120B and DeepSeek V3.1 671B exhibit relatively higher \texttt{pass} rates, suggesting stronger capabilities in producing semantically correct tests. In contrast, Devstral-Small and Gemini 2.5 Flash suffer from excessive execution-level failures (up to 80\%).

Among the evaluated models, Qwen3 Coder 480B A35B demonstrates the highest frequency of introducing unnecessary introductory text into its output. 

A notable issue observed in Go-generated tests was improper import management, specifically unused imports, along with frequent occurrences of undefined names and function references.

Python tests generated by the Devstral-Small model exhibited the broadest spectrum of exception types, encompassing ValueError, NotImplementedError, TypeError, and others.

Finally, for Java-generated tests, the most prevalent exceptions encountered were NullPointerException and IndexOutOfBoundsException.

In general, all models have errors for the same files. Only several samples were misgenerated by one or two models. It points out that all models have similar weaknesses. The Go generated tests have the highest number of errors that can be caused by the dominant amount of data in Java and Python for the models training.

Interestingly, syntax error rates remain low, implying that most models have largely mastered surface-level code formatting. Overall, setup of correct and executable test suite emerges as the principal bottleneck in achieving reliable test maintenance.

\section{Conclusion}
\label{sec:conclusion}
This work presents TAM-Eval, a novel benchmark and evaluation framework for assessing large language models (LLMs) in automated unit test maintenance across Python, Java, and Go. The benchmark comprises 1,539 task instances, carefully curated through a multi-stage process to reflect practical software engineering scenarios, with a modular design that supports future extensions to additional languages and repositories. 

The TAM-Eval framework employs reference-free metrics -- test suite pass rates, test and mutation coverage gains -- to evaluate abilities of creation, repairing, and updating unit tests, revealing significant insights into model performance. 

Our baseline analysis reveals significant shortcomings in LLM performance for unit test maintenance. 
Although performance improves with multiple attempts 
underscoring the importance of iterative feedback from automated verifier systems such as compilers.

These findings highlight the potential of LLMs in test maintenance while identifying key areas for improvement, including enhanced context handling and support for higher-order testing paradigms, paving the way for future research in automated software engineering. Beyond the experiments reported here, the TAM-Eval benchmark has also been used as a validation set for training reward models~\cite{bruches2026rmrf}, demonstrating its applicability for related downstream tasks.

\bibliographystyle{IEEEtran}
\bibliography{biblinks}

@misc{gu2025testartimprovingllmbasedunit,
      title={TestART: Improving LLM-based Unit Testing via Co-evolution of Automated Generation and Repair Iteration}, 
      author={Siqi Gu and Quanjun Zhang and others},
      year={2025},
      eprint={2408.03095},
      archivePrefix={arXiv},
      primaryClass={cs.SE},
      url={https://arxiv.org/abs/2408.03095}, 
}

@inproceedings{10.1145/3663529.3663801,
author = {Chen, Yinghao and Hu, Zehao and others},
title = {ChatUniTest: A Framework for LLM-Based Test Generation},
year = {2024},
isbn = {9798400706585},
publisher = {Association for Computing Machinery},
address = {New York, NY, USA},
url = {https://doi.org/10.1145/3663529.3663801},
doi = {10.1145/3663529.3663801},
booktitle = {Companion Proceedings of the 32nd ACM International Conference on the Foundations of Software Engineering},
pages = {572–576},
numpages = {5},
location = {Porto de Galinhas, Brazil},
series = {FSE 2024}
}

@article{10.1145/3660783,
author = {Yuan, Zhiqiang and Liu, Mingwei and others},
title = {Evaluating and Improving ChatGPT for Unit Test Generation},
year = {2024},
issue_date = {July 2024},
publisher = {Association for Computing Machinery},
address = {New York, NY, USA},
volume = {1},
number = {FSE},
url = {https://doi.org/10.1145/3660783},
doi = {10.1145/3660783},
journal = {Proc. ACM Softw. Eng.},
month = jul,
articleno = {76},
numpages = {24}
}

@misc{gu2025llmtestgenerationiterative,
      title={LLM Test Generation via Iterative Hybrid Program Analysis}, 
      author={Sijia Gu and Noor Nashid and Ali Mesbah},
      year={2025},
      eprint={2503.13580},
      archivePrefix={arXiv},
      primaryClass={cs.SE},
      url={https://arxiv.org/abs/2503.13580}, 
}

@ARTICLE{10329992,
  author={Schäfer, Max and Nadi, Sarah and others},
  journal={IEEE Transactions on Software Engineering}, 
  title={An Empirical Evaluation of Using Large Language Models for Automated Unit Test Generation}, 
  year={2024},
  volume={50},
  number={1},
  pages={85-105},
  doi={10.1109/TSE.2023.3334955}}

@inproceedings{NEURIPS2024_94f093b4,
 author = {M\"{u}ndler, Niels and M\"{u}ller, Mark Niklas and others},
 booktitle = {Advances in Neural Information Processing Systems},
 editor = {A. Globerson and L. Mackey and others},
 pages = {81857--81887},
 publisher = {Curran Associates, Inc.},
 title = {SWT-Bench: Testing and Validating Real-World Bug-Fixes with Code Agents},
 volume = {37},
 year = {2024}
}

@inproceedings{10.1145/3663529.3663839,
author = {Alshahwan, Nadia and Chheda, Jubin and others},
title = {Automated Unit Test Improvement using Large Language Models at Meta},
year = {2024},
isbn = {9798400706585},
publisher = {Association for Computing Machinery},
address = {New York, NY, USA},
url = {https://doi.org/10.1145/3663529.3663839},
doi = {10.1145/3663529.3663839},
booktitle = {Companion Proceedings of the 32nd ACM International Conference on the Foundations of Software Engineering},
pages = {185–196},
numpages = {12},
location = {Porto de Galinhas, Brazil},
series = {FSE 2024}
}

@article{10.1109/TSE.1976.233818,
author = {Miller, W. and Spooner, D. L.},
title = {Automatic Generation of Floating-Point Test Data},
year = {1976},
issue_date = {May 1976},
publisher = {IEEE Press},
volume = {2},
number = {3},
issn = {0098-5589},
url = {https://doi.org/10.1109/TSE.1976.233818},
doi = {10.1109/TSE.1976.233818},
journal = {IEEE Trans. Softw. Eng.},
month = may,
pages = {223–226},
numpages = {4}
}

@inproceedings{10.1145/1297846.1297902,
author = {Pacheco, Carlos and Ernst, Michael D.},
title = {Randoop: feedback-directed random testing for Java},
year = {2007},
isbn = {9781595938657},
publisher = {Association for Computing Machinery},
address = {New York, NY, USA},
url = {https://doi.org/10.1145/1297846.1297902},
doi = {10.1145/1297846.1297902},
booktitle = {Companion to the 22nd ACM SIGPLAN Conference on Object-Oriented Programming Systems and Applications Companion},
pages = {815–816},
numpages = {2},
location = {Montreal, Quebec, Canada},
series = {OOPSLA '07}
}

@inproceedings{10.1145/2025113.2025179,
author = {Fraser, Gordon and Arcuri, Andrea},
title = {EvoSuite: automatic test suite generation for object-oriented software},
year = {2011},
isbn = {9781450304436},
publisher = {Association for Computing Machinery},
address = {New York, NY, USA},
url = {https://doi.org/10.1145/2025113.2025179},
doi = {10.1145/2025113.2025179},
booktitle = {Proceedings of the 19th ACM SIGSOFT Symposium and the 13th European Conference on Foundations of Software Engineering},
pages = {416–419},
numpages = {4},
location = {Szeged, Hungary},
series = {ESEC/FSE '11}
}

@misc{gao2025promptalchemistautomatedllmtailored,
      title={The Prompt Alchemist: Automated LLM-Tailored Prompt Optimization for Test Case Generation}, 
      author={Shuzheng Gao and Chaozheng Wang and others},
      year={2025},
      eprint={2501.01329},
      archivePrefix={arXiv},
      primaryClass={cs.SE},
      url={https://arxiv.org/abs/2501.01329}, 
}

@misc{huang2024largelanguagemodelsselfcorrect,
      title={Large Language Models Cannot Self-Correct Reasoning Yet}, 
      author={Jie Huang and Xinyun Chen and others},
      year={2024},
      eprint={2310.01798},
      archivePrefix={arXiv},
      primaryClass={cs.CL},
      url={https://arxiv.org/abs/2310.01798}, 
}

@inproceedings{olausson2024repair,
	title        = {Is Self-Repair a Silver Bullet for Code Generation?},
	author       = {Theo X. Olausson and Jeevana Priya Inala and Chenglong Wang and Jianfeng Gao and Armando Solar-Lezama},
	year         = 2024,
	booktitle    = {International Conference on Learning Representations (ICLR)}
}

@inproceedings{10.5555/3600270.3602070,
author = {Wei, Jason and Wang, Xuezhi and others},
title = {Chain-of-thought prompting elicits reasoning in large language models},
year = {2022},
isbn = {9781713871088},
publisher = {Curran Associates Inc.},
address = {Red Hook, NY, USA},
booktitle = {Proceedings of the 36th International Conference on Neural Information Processing Systems},
articleno = {1800},
numpages = {14},
location = {New Orleans, LA, USA},
series = {NIPS '22}
}

@inproceedings{10.5555/3666122.3666639,
author = {Yao, Shunyu and Yu, Dian and others},
title = {Tree of thoughts: deliberate problem solving with large language models},
year = {2023},
publisher = {Curran Associates Inc.},
address = {Red Hook, NY, USA},
booktitle = {Proceedings of the 37th International Conference on Neural Information Processing Systems},
articleno = {517},
numpages = {14},
location = {New Orleans, LA, USA},
series = {NIPS '23}
}

@inproceedings{10.1109/ICSE48619.2023.00194,
author = {Kang, Sungmin and Yoon, Juyeon and Yoo, Shin},
title = {Large Language Models are Few-Shot Testers: Exploring LLM-Based General Bug Reproduction},
year = {2023},
isbn = {9781665457019},
publisher = {IEEE Press},
url = {https://doi.org/10.1109/ICSE48619.2023.00194},
doi = {10.1109/ICSE48619.2023.00194},
booktitle = {Proceedings of the 45th International Conference on Software Engineering},
pages = {2312–2323},
numpages = {12},
location = {Melbourne, Victoria, Australia},
series = {ICSE '23}
}

@inproceedings{10.1145/3508398.3511495,
author = {Challande, Alexis and David, Robin and Renault, Gu\'{e}na\"{e}l},
title = {Building a Commit-level Dataset of Real-world Vulnerabilities},
year = {2022},
isbn = {9781450392204},
publisher = {Association for Computing Machinery},
address = {New York, NY, USA},
url = {https://doi.org/10.1145/3508398.3511495},
doi = {10.1145/3508398.3511495},
booktitle = {Proceedings of the Twelfth ACM Conference on Data and Application Security and Privacy},
pages = {101–106},
numpages = {6},
location = {Baltimore, MD, USA},
series = {CODASPY '22}
}

@misc{bogomolov2024longcodearenaset,
      title={Long Code Arena: a Set of Benchmarks for Long-Context Code Models}, 
      author={Egor Bogomolov and Aleksandra Eliseeva and others},
      year={2024},
      eprint={2406.11612},
      archivePrefix={arXiv},
      primaryClass={cs.LG},
      url={https://arxiv.org/abs/2406.11612}, 
}

@misc{yang2025swesmith,
  title={SWE-smith: Scaling Data for Software Engineering Agents}, 
  author={John Yang and Kilian Leret and others},
  year={2025},
  eprint={2504.21798},
  archivePrefix={arXiv},
  primaryClass={cs.SE},
  url={https://arxiv.org/abs/2504.21798}, 
}

@inproceedings{
    jimenez2024swebench,
    title={{SWE}-bench: Can Language Models Resolve Real-world Github Issues?},
    author={Carlos E Jimenez and John Yang and Alexander Wettig and Shunyu Yao and Kexin Pei and Ofir Press and Karthik R Narasimhan},
    booktitle={The Twelfth International Conference on Learning Representations},
    year={2024},
    url={https://openreview.net/forum?id=VTF8yNQM66}
}

@misc{xu2025clovertestcasegeneration,
      title={CLOVER: A Test Case Generation Benchmark with Coverage, Long-Context, and Verification}, 
      author={Jiacheng Xu and Bo Pang and Jin Qu and Hiroaki Hayashi and Caiming Xiong and Yingbo Zhou},
      year={2025},
      eprint={2502.08806},
      archivePrefix={arXiv},
      primaryClass={cs.SE},
      url={https://arxiv.org/abs/2502.08806}, 
}

@misc{wang2025projecttestprojectlevelllmunit,
      title={ProjectTest: A Project-level LLM Unit Test Generation Benchmark and Impact of Error Fixing Mechanisms}, 
      author={Yibo Wang and Congying Xia and others},
      year={2025},
      eprint={2502.06556},
      archivePrefix={arXiv},
      primaryClass={cs.SE},
      url={https://arxiv.org/abs/2502.06556}, 
}

@misc{bhargava2024cpputbenchllmswritecomplex,
      title={CPP-UT-Bench: Can LLMs Write Complex Unit Tests in C++?}, 
      author={Vaishnavi Bhargava and Rajat Ghosh and Debojyoti Dutta},
      year={2024},
      eprint={2412.02735},
      archivePrefix={arXiv},
      primaryClass={cs.SE},
      url={https://arxiv.org/abs/2412.02735}, 
}

@article{zhang2024testbench,
  title={TestBench: Evaluating Class-Level Test Case Generation Capability of Large Language Models},
  author={Zhang, Quanjun and Shang, Ye and others},
  journal={arXiv preprint arXiv:2409.17561},
  year={2024}
}

@inproceedings{wang-etal-2025-testeval,
    title = "{TESTEVAL}: Benchmarking Large Language Models for Test Case Generation",
    author = "Wang, Wenhan  and
      Yang, Chenyuan  and
      others",
    editor = "Chiruzzo, Luis  and
      Ritter, Alan  and
      Wang, Lu",
    booktitle = "Findings of the Association for Computational Linguistics: NAACL 2025",
    month = apr,
    year = "2025",
    address = "Albuquerque, New Mexico",
    publisher = "Association for Computational Linguistics",
    url = "https://aclanthology.org/2025.findings-naacl.197/",
    pages = "3547--3562",
    ISBN = "979-8-89176-195-7"
}

@misc{yang2025qwen3technicalreport,
      title={Qwen3 Technical Report}, 
      author={An Yang and Anfeng Li and others},
      year={2025},
      eprint={2505.09388},
      archivePrefix={arXiv},
      primaryClass={cs.CL},
      url={https://arxiv.org/abs/2505.09388}, 
}

@misc{mistral2025devstral,
  author       = {{Mistral AI}},
  title        = {{DevStral: Introducing the best open-source model for coding agents.}},
  howpublished = {\url{https://mistral.ai/news/devstral}},
  note         = {Accessed: 2025-05-30},
  year         = {2025}
}

@misc{openai2025gpt41,
  author       = {{OpenAI}},
  title        = {{Introducing GPT-5 in the API}},
  howpublished = {\url{https://openai.com/index/introducing-gpt-5/}},
  note         = {Accessed: 2025-10-28},
  year         = {2025}
}

@misc{deepseekai2024deepseekv3technicalreport,
      title={DeepSeek-V3 Technical Report}, 
      author={DeepSeek-AI},
      year={2024},
      eprint={2412.19437},
      archivePrefix={arXiv},
      primaryClass={cs.CL},
      url={https://arxiv.org/abs/2412.19437}, 
}

@misc{jiang2024surveylargelanguagemodels,
      title={A Survey on Large Language Models for Code Generation}, 
      author={Juyong Jiang and Fan Wang and others},
      year={2024},
      eprint={2406.00515},
      archivePrefix={arXiv},
      primaryClass={cs.CL},
      url={https://arxiv.org/abs/2406.00515}, 
}

@misc{yang2024evaluationlargelanguagemodels,
      title={On the Evaluation of Large Language Models in Unit Test Generation}, 
      author={Lin Yang and Chen Yang and others},
      year={2024},
      eprint={2406.18181},
      archivePrefix={arXiv},
      primaryClass={cs.SE},
      url={https://arxiv.org/abs/2406.18181}, 
}

@inproceedings{dvivedi2024analysis,
author = {Dvivedi, Shubhang Shekhar and Vijay, Vyshnav and others},
title = {A Comparative Analysis of Large Language Models for Code Documentation Generation},
year = {2024},
isbn = {9798400706851},
publisher = {Association for Computing Machinery},
address = {New York, NY, USA},
url = {https://doi.org/10.1145/3664646.3664765},
doi = {10.1145/3664646.3664765},
booktitle = {Proceedings of the 1st ACM International Conference on AI-Powered Software},
pages = {65–73},
numpages = {9},
location = {Porto de Galinhas, Brazil},
series = {AIware 2024}
}

@misc{cordeiro2024empiricalstudycoderefactoring,
      title={An Empirical Study on the Code Refactoring Capability of Large Language Models}, 
      author={Jonathan Cordeiro and Shayan Noei and Ying Zou},
      year={2024},
      eprint={2411.02320},
      archivePrefix={arXiv},
      primaryClass={cs.SE},
      url={https://arxiv.org/abs/2411.02320}, 
}

@misc{meng2024empiricalstudyllmbasedagents,
      title={An Empirical Study on LLM-based Agents for Automated Bug Fixing}, 
      author={Xiangxin Meng and Zexiong Ma and others},
      year={2024},
      eprint={2411.10213},
      archivePrefix={arXiv},
      primaryClass={cs.SE},
      url={https://arxiv.org/abs/2411.10213}, 
}

@book{10.5555/1349795,
author = {Huizinga, Dorota and Kolawa, Adam},
title = {Automated Defect Prevention: Best Practices in Software Management},
year = {2007},
isbn = {0470042125},
publisher={Wiley-IEEE Computer Society Pr}
}

@misc{globalapptestingMuchDoes,
	author = {{Global App Testing}},
	title = {{H}ow {M}uch {D}oes {S}oftware {T}esting {C}ost in 2025? --- globalapptesting.com},
	howpublished = {\url{https://www.globalapptesting.com/blog/software-testing-cost}},
	year = {},
	note = {[Accessed 26-05-2025]},
}

@misc{li2024promptinglargelanguagemodels,
      title={Prompting Large Language Models to Tackle the Full Software Development Lifecycle: A Case Study}, 
      author={Bowen Li and Wenhan Wu and others},
      year={2024},
      eprint={2403.08604},
      archivePrefix={arXiv},
      primaryClass={cs.CL},
      url={https://arxiv.org/abs/2403.08604}, 
}

@misc{jain2025testgenevalrealworldunit,
      title={TestGenEval: A Real World Unit Test Generation and Test Completion Benchmark}, 
      author={Kush Jain and Gabriel Synnaeve and Baptiste Rozière},
      year={2025},
      eprint={2410.00752},
      archivePrefix={arXiv},
      primaryClass={cs.SE},
      url={https://arxiv.org/abs/2410.00752}, 
}

@misc{openai2025gptoss120bgptoss20bmodel,
      title={GPT-OSS-120b \& GPT-OSS-20b Model Card}, 
      author={OpenAI},
      year={2025},
      eprint={2508.10925},
      archivePrefix={arXiv},
      primaryClass={cs.CL},
      url={https://arxiv.org/abs/2508.10925}, 
}

@ARTICLE{2025arXiv250706261C,
       author = {{Comanici}, Gheorghe and {Bieber}, Eric and others},
        title = {Gemini 2.5: Pushing the Frontier with Advanced Reasoning, Multimodality, Long Context, and Next Generation Agentic Capabilities},
      journal = {arXiv e-prints},
         year = 2025,
        month = jul,
          eid = {arXiv:2507.06261},
        pages = {arXiv:2507.06261},
          doi = {10.48550/arXiv.2507.06261},
archivePrefix = {arXiv},
       eprint = {2507.06261},
 primaryClass = {cs.CL},
       adsurl = {https://ui.adsabs.harvard.edu/abs/2025arXiv250706261C}
}

@misc{chervyakov2025meracodeunifiedframework,
      title={MERA Code: A Unified Framework for Evaluating Code Generation Across Tasks}, 
      author={Artem Chervyakov and Alexander Kharitonov and others},
      year={2025},
      eprint={2507.12284},
      archivePrefix={arXiv},
      primaryClass={cs.SE},
      url={https://arxiv.org/abs/2507.12284}, 
}

@inproceedings{bruches2026rmrf,
  author    = {Elena Bruches and Daniil Grebenkin and others},
  title     = {{RM -RF}: Reward Model for Run-Free Unit Test Evaluation},
  booktitle = {Proceedings of the 33rd IEEE International Conference on Software Analysis, Evolution and Reengineering (SANER)},
  year      = {2026},
  note      = {To be published},
  url       = {https://arxiv.org/abs/2601.13097}
}

\end{document}